\documentstyle[twoside,twocolumn,prl,aps,epsf]{revtex}
\input{epsf}

\begin{document}
\input{psfig}
\twocolumn[\hsize\textwidth\columnwidth\hsize\csname @twocolumnfalse\endcsname 
\title{Breakdown of scale-invariance in the coarsening of phase-separating
binary fluids}
\author{Alexander J. Wagner and J.M. Yeomans}
\address{Theoretical Physics, Oxford University,
1 Keble Rd. Oxford OX1 3NP, UK}
\date{\today}
\maketitle
\begin{abstract}
We present evidence, based on lattice Boltzmann simulations, to show
that the coarsening of the domains in phase separating binary fluids
is not a scale-invariant process. Moreover we emphasise that the pathway
by which phase separation occurs depends strongly on the
relation between diffusive and hydrodynamic time scales. 

\vspace{-0.9cm}
\pacs P PACS numbers 47.20Hw,  47.55K,  05.70Ln.

\vspace{0.9cm}
\end{abstract}
]

Our aim in this Letter is to discuss domain growth in
two-dimensional binary fluid mixtures. 
When a binary mixture of fluids, $A$ and $B$ say, is quenched below
its critical point it phase separates into an $A$-rich and a $B$-rich
phase. 
This is achieved by domains of
the two phases forming and then growing.
It has been widely assumed that the coarsening of the domains 
is a scale-invariant process \onlinecite{B94}. We present
evidence, based on lattice Boltzmann simulations \onlinecite{OS95}, 
to show that 
this is not the case for binary fluids where
different growth mechanisms compete at all times. We also emphasise 
that the pathways by
which coarsening occurs have a strong  qualitative
dependence on the
relation between diffusive and hydrodynamic time scales.

There is a large body of numerical and experimental data on domain
coarsening  in systems without hydrodynamics, such as magnets or
binary alloys. The conclusion is that, after initial transients, the
domain growth is scale invariant. The
morphology of the domain pattern remains statistically equivalent
at all times apart from a change of the length scale and the correlation
function of the order parameter $\phi$ obeys the scaling form
\begin{equation}
<\phi({\bf x+r},t) \phi({\bf x},t)>=f({\bf r}/{R(t)})
\end{equation}
where $<\dots >$ indicates a spatial average and $R(t)$ is a length
scale which is typically observed to grow as a power law
\begin{equation}
R(t) \sim (t-t_0)^{\alpha}.
\end{equation}
$\alpha$, the growth exponent, 
is a universal constant which is expected to depend only on the growth
mechanism and not on the microscopic details of the system.
$t_0$ is a zero time that
does not have to coincide with the start of the simulation. 

Several different growth mechanisms have been identified or proposed
for bicontinuous two-dimensional binary fluids.

\noindent {\bf Diffusive growth:} this is the mechanism by which
domains grow by the diffusion of material between
them\onlinecite{LS61}. The growth is slow with $\alpha =1/3$ as
material has to diffuse across a $B$ domain to move between $A$
domains and vice versa. This mechanism is common to all materials with
a conserved order parameter (e.g. binary alloys, spin systems with
Kawasaki dynamics) and does not rely on the hydrodynamic properties of
fluids.

\noindent {\bf Diffusion-enhanced collisions:} In a system of
concentrated droplets the diffusion field around the droplets leads to
an attraction between them\onlinecite{T95}.  In a fluid they are
able to flow in response leading to a faster coalescence, but with
$\alpha$ still $1/3$.

\noindent {\bf Hydrodynamic growth:} for time scales
over which hydrodynamic modes can be excited bulk fluid flow is
possible\onlinecite{S79}. This is a faster process which leads to a
growth exponent $\alpha=2/3$ for fluids at long times. However, as we
shall demonstrate below, hydrodynamic flow, driven by the pressure
difference between points of different curvature, is effective in
reducing the interface length and making domains more nearly circular
but not in enhancing coalescence of domains.

\noindent {\bf Noise-induced growth:} Noise can lead to a growth
exponent $\alpha=1/2$ \onlinecite{SG85}.  However there is no noise
in the lattice Boltzmann results presented here and this mechanism
will not be relevant.

In this letter we present evidence that:
1. The competition between diffusive and hydrodynamic growth leads
to a breakdown of scale invariance.
2. For very low viscosities capillary waves are important and lead
to yet another possible growth mechanism.
3. The relative magnitudes of the diffusion constant, viscosity and
surface tension are qualitatively important in determining the
route along which domain growth proceeds. In particular we clearly identify the
double quench pathway first described by Tanaka\onlinecite{T00}.

Our evidence for the breakdown of scale invariance in hydrodynamic
systems is encapsulated in Figures 1,2, and 3 which correspond to
high, intermediate and low viscosities respectively. The left-hand
column of snapshot pictures in each figure shows the domain pattern at
three times. Coarsening of the domains can be clearly seen and it is
immediately apparent that the shapes of the growing domains depend on the
value of the viscosity.

To more easily compare the emerging patterns the right-hand column of
snapshots in each of the figures shows an enlargement of part of the
corresponding left-hand picture. The enlargement is by a factor
$(t_3/t)^{\alpha^1}$ for a picture at time $t$ where $t_3$ is the time
of the final snap-

\begin{figure}[t]
\begin{center}
\begin{minipage}{6cm}
\vspace{0.7cm}
\end{minipage}\\
\begin{minipage}[t]{4cm}
\centerline{\psfig{figure=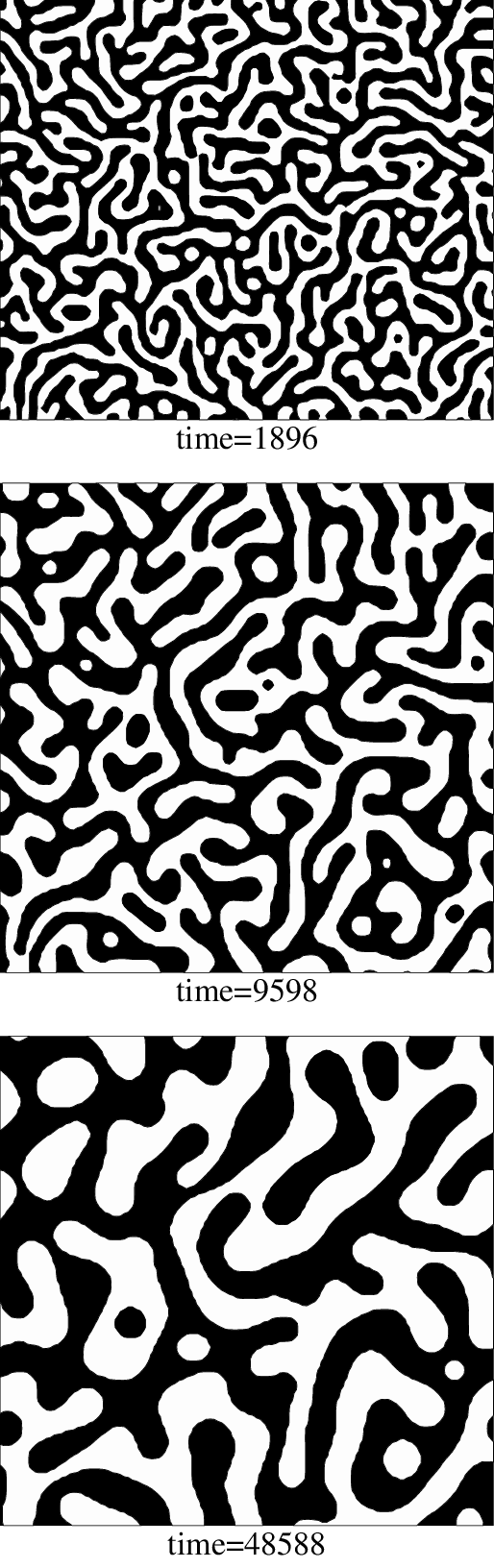,width=4cm}}
\end{minipage}
\begin{minipage}[t]{4cm}
\centerline{\psfig{figure=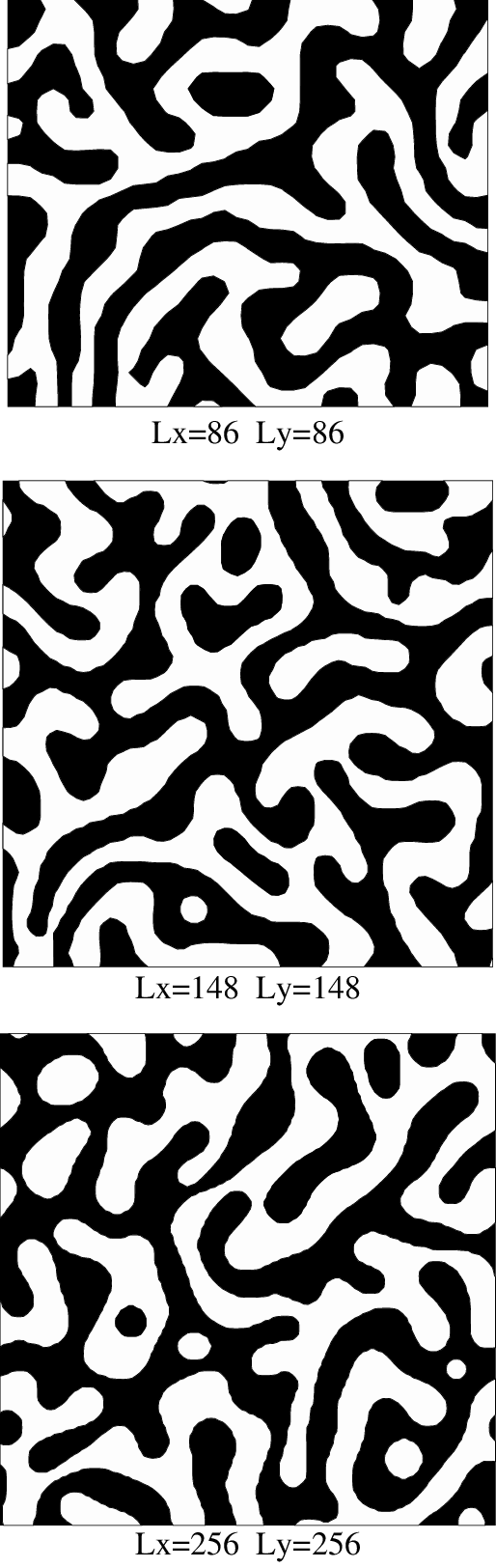,width=4cm}}
\end{minipage}
\begin{minipage}[t]{6cm}
\centerline{\psfig{figure=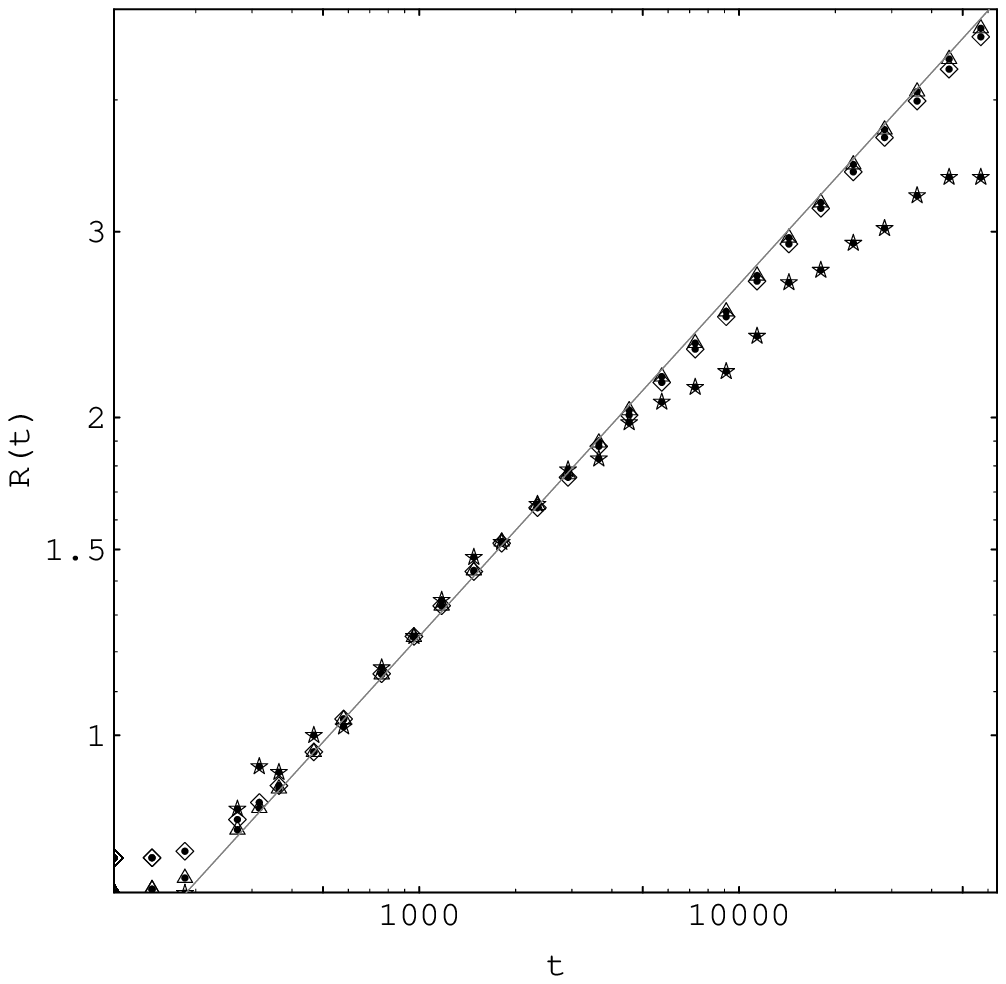,width=5.5cm}}
\end{minipage}
\end{center}
\caption{Evolution of phase-separating domains for a binary fluid with
high viscosity. The left-hand column shows the phase ordering process
at three different times. The right-hand column shows each snapshot
scaled by a factor $(t/48588)^\frac{1}{3}$. The resulting box size
(Lx,Ly) is indicated. The graph at the base of the figure shows the
behaviour of $\ln R$ as a function of $\ln t$ where $R$ is a length
scale and $t$ the time. Different scales are distinguished by
$\diamond (R^1)$, $\triangle (R^{\circ})$, and $* (R^{\#})$. The line
corresponds to  $\alpha=1/3$.}
\label{fig:highvisc}
\end{figure} 

\noindent shot and $\alpha^1$ is the growth exponent for the
length scale $R^1$ defined below. If the system is scale invariant all
the figures in the right-hand column would be expected to look
(statistically) identical.

To obtain a more quantitative measure of the growth a graph of $\ln
R(t)$ as a function of $\ln t$ is also displayed for each value of the
viscosity. Three different measures of length are considered.
Previous work\onlinecite{OO95} has concentrated
on measuring $R^{1}(t)$, the inverse first moment of the circularly
averaged structure factor.
Here we also present results for length scales derived from the  
length of the interface $L_I$ and 
and the number of domains $N$ 
\begin{equation}
R^{\circ}(t)=\frac{L_x*L_y}{L_I} ,\;\;\;\;\;\;\;
R^{\#}(t)=\sqrt{\frac{L_x L_y}{N}}\label{number}
\end{equation}
respectively where $L_x$ and $L_y$ are the linear dimensions of the
simulation box.

If the system is scale invariant all lengths should
scale with the same value of $\alpha$. (This excludes microscopic
lengths, such as the lattice spacing and interface widths, which remain
unchanged during the growth process.) 
We show that the different lengths do not scale in the same way for some
values of the fluid viscosity. This implies a breakdown of scale
invariance which is further supported by visual inspection of the
growing domains.

We consider each value of the viscosity in turn. At very high viscosities
(Figure 1) hydrodynamics is unimportant. Therefore the system
resembles Model B (growth with a conserved order parameter) in the
language of critical phenomena. Diffusive growth, with $\alpha=1/3$,
is expected. The length scales $R^1$ and $R^{\circ}$ clearly show this
behaviour. For early times $R^{\#}$ also scales as $t^{1/3}$ but there
is a slower growth for later times. This was a feature of all the
simulations we ran and became more pronounced for the larger
systems. A possible explanation lies in the lack of a bicontinuous
structure in two dimensions which leads to  the formation of nested
structures which affect the growth.
The snapshots on the right-hand side of Figure 1
show no visual evidence for a breakdown of scaling.

Figure 2 shows results for intermediate viscosities which are low
enough to allow hydrodynamic flow but sufficiently high to damp out
capillary waves. The well-known hydrodynamic growth exponent $2/3$ is
observed for $R^{1}$ and $R^{\circ}$ but for $R^{\#}$ the growth
exponent quickly crosses over to $\alpha^\#=1/3$. This implies that
the number of domains is decreasing more slowly than in a
scale-invariant state. It occurs because, although the interface
curvature of the domains is rapidly decreased by the flow, once the
domains are circular hydrodynamics can only assist the growth through
the much slower diffusion-enhanced collisions which proceed with
$\alpha=1/3$\onlinecite{T95}.

As a result domains on large length scales are tortuous. The smaller
the domains the more circular they become.  The length scale of the
crossover between these behaviours increases with time. Hence as time
progresses an increasingly deep hierarchy of circular domains within

\begin{figure}[t]
\begin{center}
\begin{minipage}[t]{8cm}
\centerline{\psfig{figure=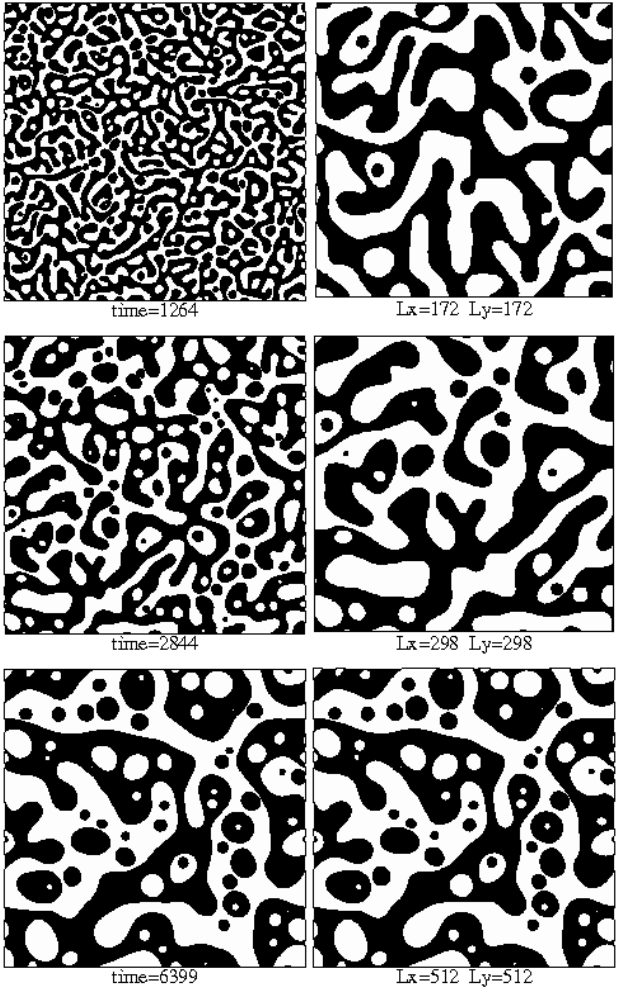,width=8.3cm}}
\end{minipage}
\begin{minipage}[t]{6cm}
\centerline{\psfig{figure=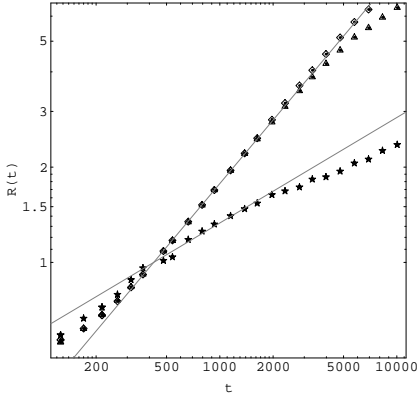,width=5.5cm}}
\end{minipage}
\end{center}
\caption{As Figure 1 but for an intermediate value of the
viscosity. Scaling is by a factor $(t/6399)^\frac{2}{3}$. The lines correspond to $\alpha=2/3$ and
$1/3$.}
\label{fig:medvisc}
\end{figure} 

\noindent circular domains results. Pictorial evidence for the lack of
scale invariance in the growth process can be seen in the scaled snapshots
in Figure 2.

In most simulations of hydrodynamic growth results have been
limited to $R^1$ (but see \onlinecite{menke} where several different
measures are used for a one-component fluid). This gives most weight
to the largest domains which continue 

\begin{figure}[t]
\begin{center}
\begin{minipage}[t]{8cm}
\centerline{\psfig{figure=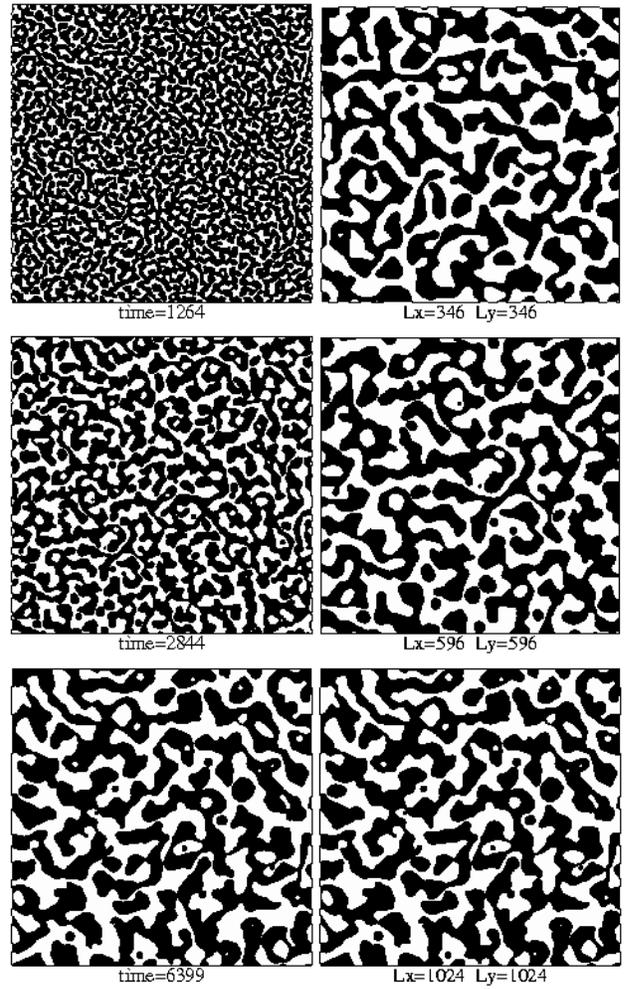,width=8.3cm}}
\end{minipage}
\begin{minipage}[t]{6cm}
\centerline{\psfig{figure=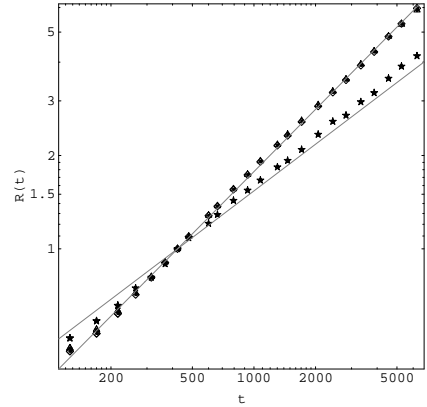,width=5.5cm}}
\end{minipage}
\end{center}
\caption{As Figure 1 but for a low value of the viscosity. Scaling is
by a factor $(t/6399)^\frac{2}{3}$. The lines
correspond to $\alpha=2/3$ and $1/2$.}
\label{fig:lowvisc}
\end{figure} 

\noindent to grow via the $2/3$ power law and this is the
exponent that has been widely reported \onlinecite{OO95}.  The measure
$R^{\circ}$ is derived from the interface length. Initially this is
mostly in the larger domains and $R^{\circ}$ is not sensitive to the
structural change. For late times there is some indication that the
growth of this measure is slowing down. $R^{\#}$ however is related to
the number of domains and provides a good

\begin{figure}[t]
\begin{center}
\begin{minipage}{6cm}
\vspace{0.7cm}
\end{minipage}\\
\begin{minipage}[t]{4cm}
\centerline{\psfig{figure=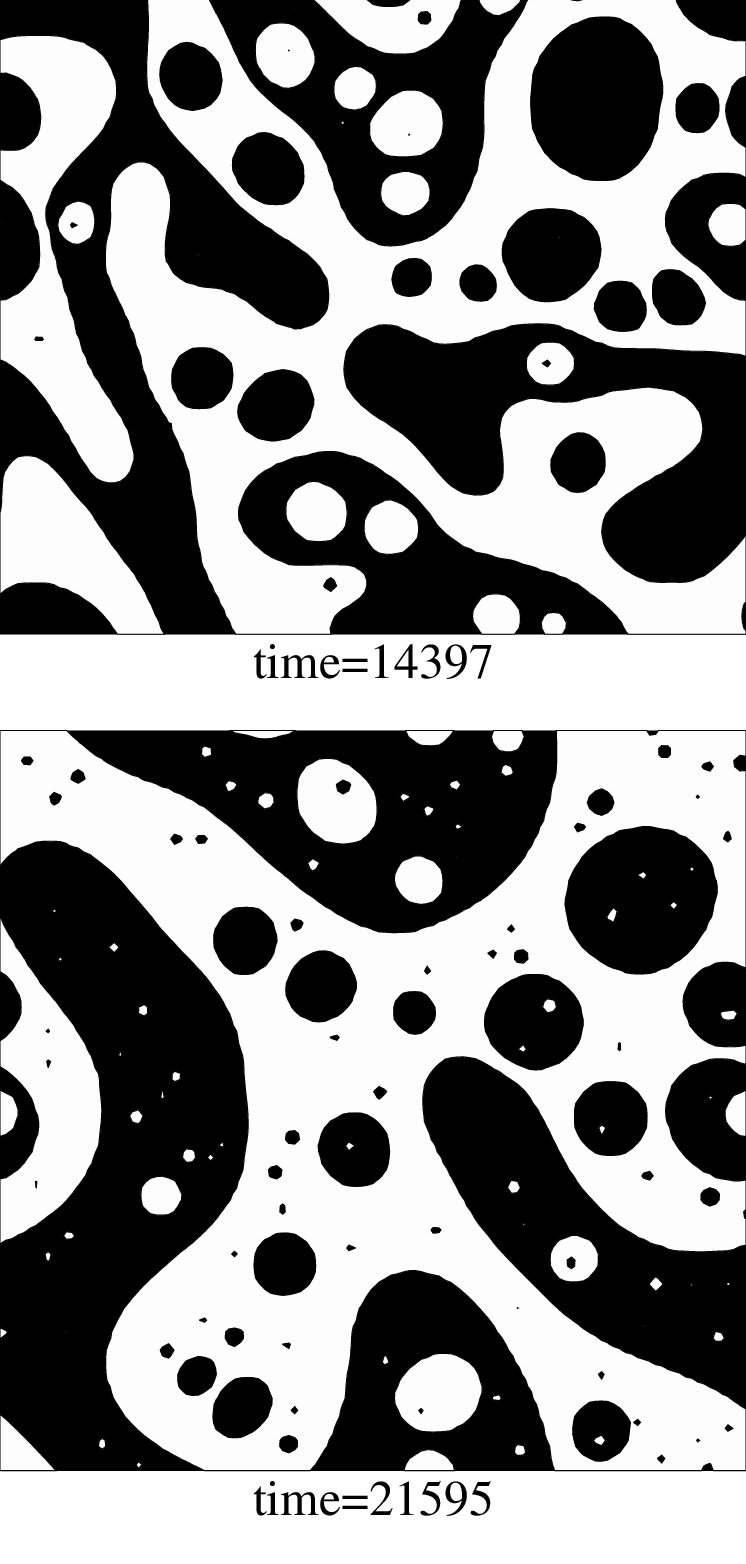,width=4cm}}
\end{minipage}
\begin{minipage}[t]{4cm}
\centerline{\psfig{figure=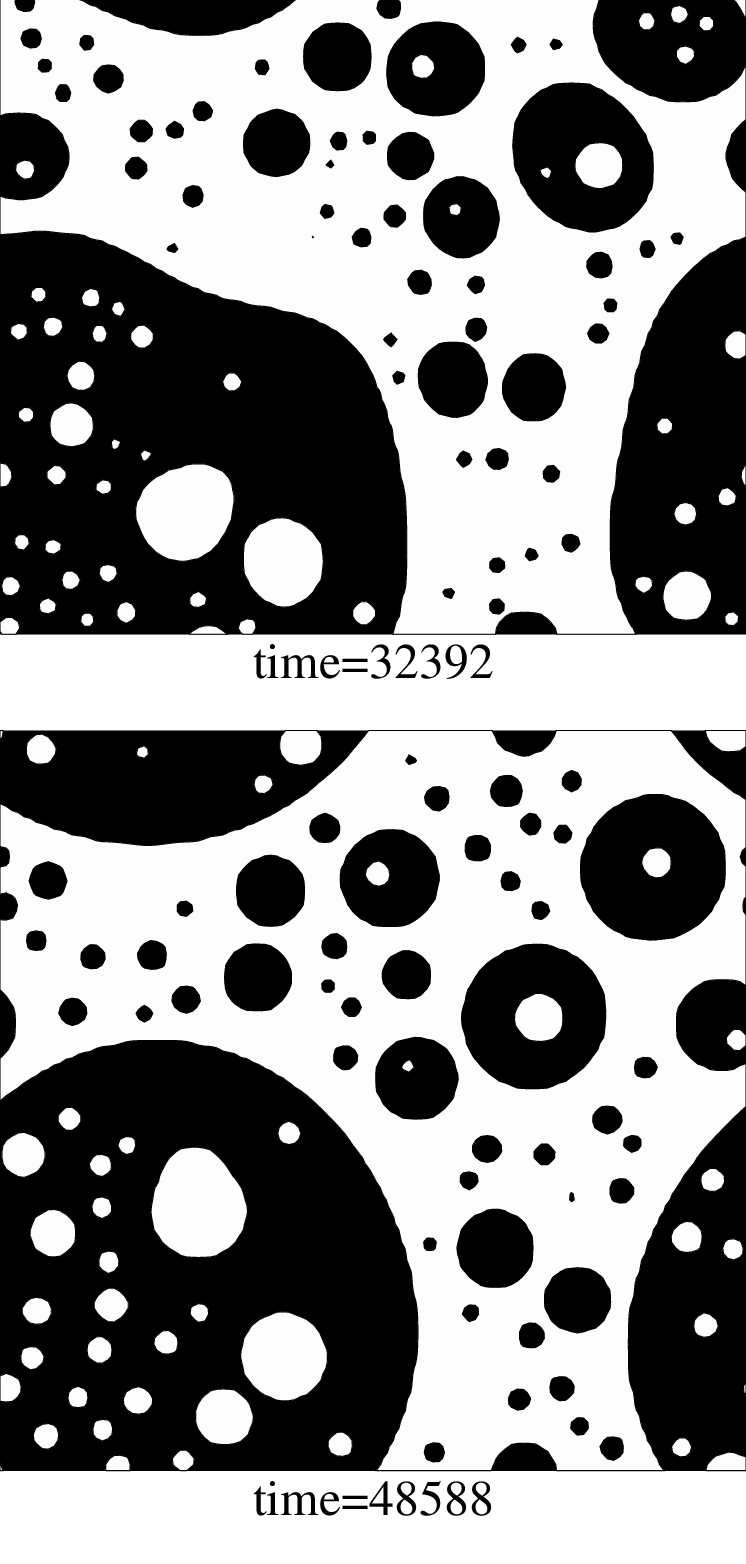,width=4cm}}
\end{minipage}\\
\begin{minipage}[t]{6cm}
\centerline{\psfig{figure=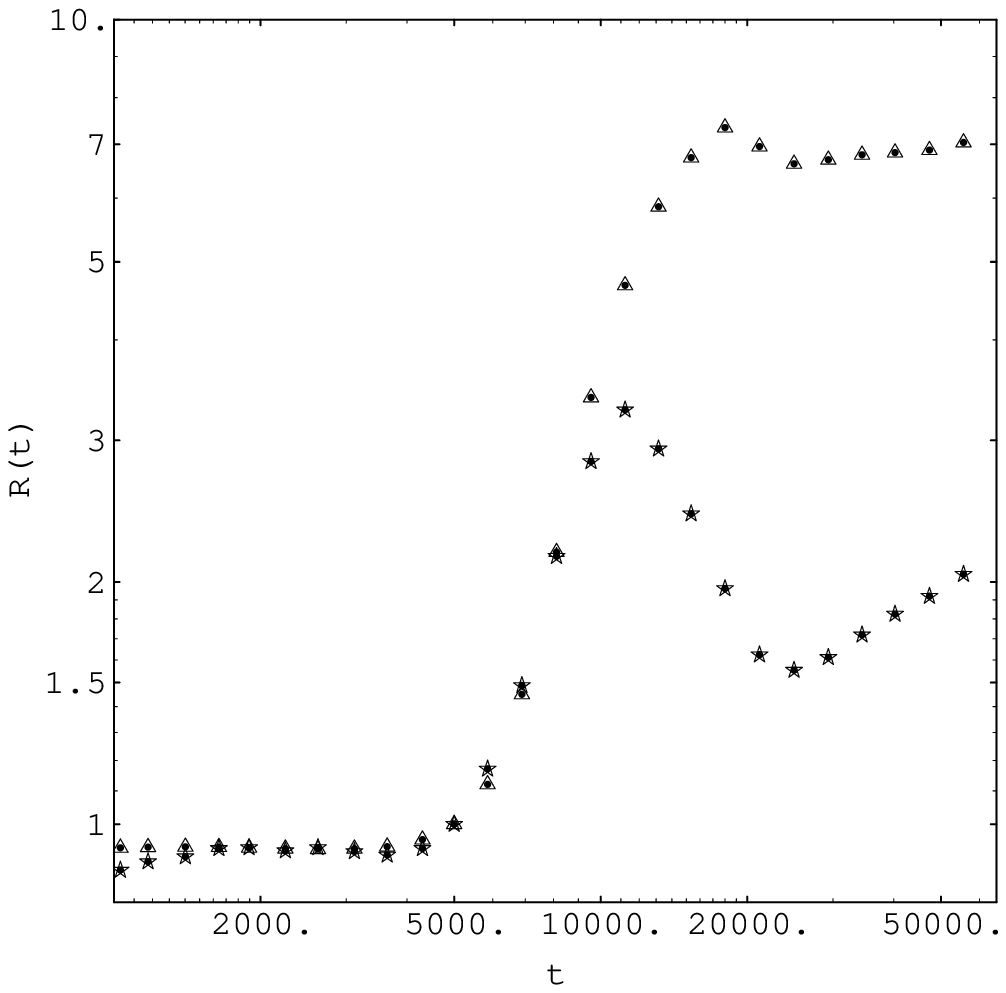,width=4.06cm}}
\end{minipage}
\end{center}
\caption{The evolution with time of phase separating domains for a
binary fluid with intermediate viscosity but low diffusivity. Domains
quickly become circular but the order parameter does not reach its
equilibrium value. Hence a second crop of domains forms by spinodal
decomposition. Tanaka has termed this a double quench. $R(t)$ shows no
clear scaling behaviour (see Figure 1 caption for definition of symbols).}
\label{fig:double}
\end{figure} 

\noindent  measure of the smaller scale
features of the pattern.

At low viscosities (Figure 3) the morphology of the domain growth is
again altered. Now the damping of capillary waves is so small that
their amplitude can be of order of the domain size. The real-space
pictures of the growth show rugged interfaces. This enhances the
domain coalescence and the growth law for $R^{\#}$ is
changed from $1/3$ to $1/2$.

Finally in Figure 4 we show yet another pathway to phase separation in
a binary fluid, the double quench first described by
Tanaka\onlinecite{T00}. These
results were obtained for intermediate viscosities but a low value of
the diffusion coefficient. Domains are created and hydrodynamic flow
allows them to attain a circular shape. However, because of the weak
diffusivity this occurs before the order parameter in each domain
attains its equilibrium value. Hence secondary spinodal decomposition
can take place within each domain and a nested hierarchy of circular
domains form up to a length scale which increases with time. Plots of
the the logarithm of the length scales against the logarithm of time in 
Figure 4 show no clear scaling behaviour on length and time scales
accessible to the simulations.

These results were obtained using the lattice Boltzmann approach
described in \onlinecite{OS95}. This gives a numerical solution of
the continuity, Navier-Stokes and convection-diffusion equations
describing the flow of binary fluids. There are two particularly
relevant advantages of the method. Firstly it is possible to vary the
diffusivity and the viscosity over a wide range. Secondly
the equilibrium state minimises an input free energy which here is
taken to be that of a model binary fluid.

To conclude, domain growth in binary fluids is a richer phenomenon
than hitherto described. The dominant growth mechanism depends
strongly on the viscosity and diffusivity of the fluid and may be
different at different length and time scales.  Many questions remain,
among them clarification of the growth mechanisms at non-symmetric
compositions, investigation of the r\^ole of noise, clarification of
the importance of self-diffusion in liquid--gas systems and assessment
of the effect of anisotropy in the dynamics of the two phases.\\
Acknowledgments: We thank G. Gonnella, E. Orlandini, B. Buck and
A. Rutenberg for enlightening discussions. JMY acknowledges support
from the EPSRC grant no GR/K97783.

\end{document}